\documentclass[runningheads]{llncs}
\usepackage[title]{appendix}
\usepackage[utf8]{inputenc}
\usepackage{graphicx}
\usepackage{ amssymb }
\usepackage{float}
\usepackage[a4paper, lmargin=2.5 cm,rmargin=2.5 cm,tmargin=2.5 cm,bmargin=2 cm]{geometry}
\usepackage{hyperref}
\usepackage{tikz-network}
\usepackage[parfill]{parskip}    
\usepackage{textcomp}
\usepackage{amsmath}
\usepackage{listings}
\usepackage{caption, subcaption}

\usepackage{enumitem}
\usepackage{xcolor}

\hypersetup{
  colorlinks=true,
  linkcolor=blue,
  citecolor=blue,
  urlcolor=blue,
  pdftitle={Malicious Activity in Online Social Networks; How Bots are Driving Discussion around the Russia/Ukraine war},
}

\date{}
\newlist{questions}{enumerate}{2}
\setlist[questions,1]{label=RQ\arabic*.,ref=RQ\arabic*}
\setlist[questions,2]{label=(\alph*),ref=\thequestionsi(\alph*)}
\begin{document}
\title{\#IStandWithPutin versus \#IStandWithUkraine: \\ The interaction of bots and humans in discussion of the Russia/Ukraine war}
\titlerunning{The interaction of bots and humans in discussion of the Russia/Ukraine war}

\author{Bridget Smart\inst{1}\orcidID{0000-0002-0910-9470}
\and
Joshua Watt\inst{1}\orcidID{0000-0001-7899-1244} \and \\
Sara Benedetti \inst{1}\orcidID{0000-0003-3514-797X}\and
Lewis Mitchell \inst{1} \orcidID{0000-0001-8191-1997}
\and \\
Matthew Roughan \inst{1} \orcidID{0000-0002-7882-7329}}
\authorrunning{B. Smart et al.}
%

\institute{\textsuperscript{1} The University of Adelaide\\
\email{\{bridget.smart, joshua.watt, sara.benedetti, lewis.mitchell, matthew.roughan\} @adelaide.edu.au}\\
\url{set.adelaide.edu.au/mathematical-sciences}}
\maketitle              
\begin{abstract}
The 2022 Russian invasion of Ukraine emphasises the role social media plays in modern-day warfare, with conflict occurring in both the physical and information environments. 
There is a large body of work on identifying malicious cyber-activity, but less focusing on the effect this activity has on the overall conversation, especially with regards to the Russia/Ukraine Conflict. 
Here, we employ a variety of techniques including information theoretic measures, sentiment and linguistic analysis, and time series techniques to understand how bot activity influences wider online discourse.
By aggregating account groups we find significant information flows from bot-like accounts to non-bot accounts with behaviour differing between sides. Pro-Russian non-bot accounts are most influential overall, with information flows to a variety of other account groups. No significant outward flows exist from pro-Ukrainian non-bot accounts, with significant flows from pro-Ukrainian bot accounts into pro-Ukrainian non-bot accounts.
We find that bot activity drives an increase in conversations surrounding angst (with $p = 2.450 \times 10^{-4}$) as well as those surrounding work/governance (with $p = 3.803 \times 10^{-18}$). 
Bot activity also shows a significant relationship with non-bot sentiment (with $p = 3.76 \times 10^{-4}$), where we find the relationship holds in both directions.
This work extends and combines existing techniques to quantify how bots are influencing people in the online conversation around the Russia/Ukraine invasion. 
It opens up avenues for researchers to understand quantitatively how these malicious campaigns operate, and what makes them impactful.

\keywords{Bot Nets \and Information Flow \and Sentiment Analysis \and Linguistic Analysis \and Disinformation Campaigns \and Influence Campaigns \and Twitter.} 

\end{abstract}

\section{Introduction}
\begin{figure}[htb!]
     \centering
    \includegraphics[width=\linewidth]{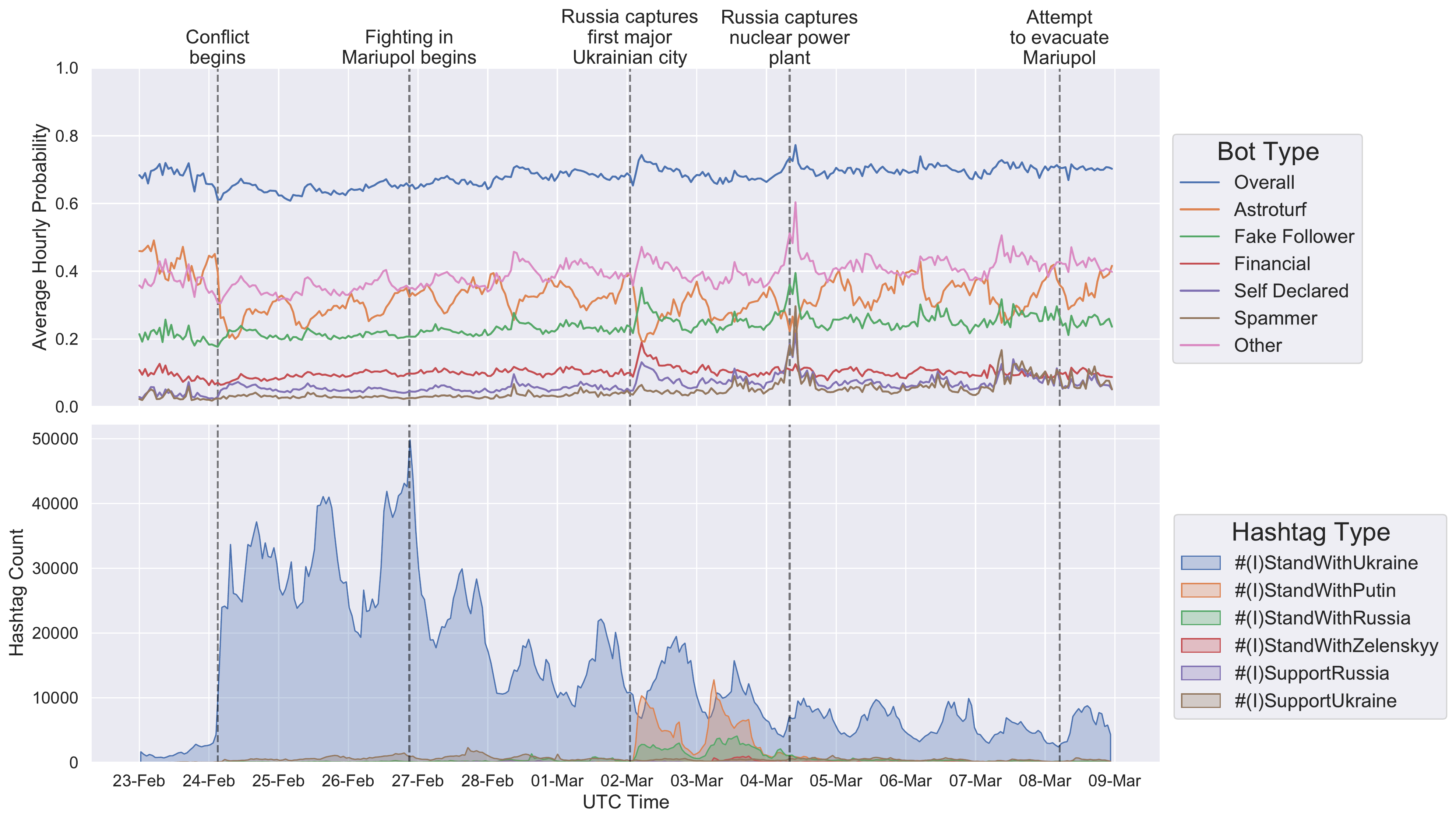}
    \captionsetup{font=footnotesize}
    \caption[0.4\textwidth]{Average hourly probabilities of bots tweeting query hashtags (top). Hourly frequency of the query hashtags (bottom). The time period we consider is the first fortnight after Russia's invasion of Ukraine. Both plots also include five significant events over this time period. Note that the query hashtags can be found in Section \ref{s:data}. We can observe a significant spike in the bot activity of several bot types on the 2nd and 4th of March. The spike in bot activity on the 2nd of March aligns with Russia's capture of Kherson, and also aligns with a significant increase in pro-Russia hashtags. This spike in activity was due to an increase in activity of pro-Russian bots -- likely used by Russian authorities. The spike in bot activity on the 4th of March aligns with when the use of pro-Russia hashtags diminished, but also when Russia captured the Zaporizhzhia nuclear power plant. This spike was due to an increase in activity of pro-Russian bots (before being removed) and an increase in activity of pro-Ukrainian bots -- likely by pro-Ukrainian authorities in response to Russian bots.\vspace{-5mm}}
    \label{fig:time_series_plots}
\end{figure}
Social media is a critical tool in information warfare, playing a large role in the 2022 Russian invasion of Ukraine~\cite{chen2022tweets,polyzos2022escalating}. 
Disinformation and more generally {\em reflexive control}~\cite{thomas2004russia} have been used by Russia and other countries against their enemies and internally for many years~\cite{doroshenko2021trollfare}. 
A relative newcomer in this space -- Twitter -- has already been extensively used for such purposes during military conflicts, for instance in Donbass \cite{doroshenko2021trollfare}, but its role in conflicts is evolving and not fully understood. 
Both sides in the Ukrainian conflict use the online information environment to influence geopolitical dynamics and sway public opinion.  
Russian social media pushes narratives around their motivation, and Ukrainian social media aims to foster and maintain external support from Western countries, as well as promote their military efforts while undermining the perception of the Russian military.  
Examples of these narratives include allegations: that Ukraine was developing biological weapons \cite{wong_us_2022}, that President Volodymyr Zelenskyy had surrendered \cite{champion_ukraines_2022,klepper_russian_2022}, and that there is a sustained campaign showing the apparent success of `The Ghost of Kiev' \cite{laurence_how_2022}. Some of the information being pushed is genuine, and some is malicious. 
It is not easy to discriminate which is which.

Understanding and measuring information flows and various language features has previously allowed researchers to understand community dynamics and identify inauthentic accounts and content \cite{south_information_2022,pondComplexContagionFeatures2020,bagrowInformationFlowReveals2019}. Here we apply and extend these techniques to understand and quantify the influence of bot-like accounts on online discussions, using Twitter data focussed on the Russian invasion of Ukraine. In essence we seek to determine whether the malicious influence campaigns work as intended.

Our dataset consists of 5,203,764 tweets, retweets, quote tweets and replies posted to Twitter between February 23rd and March 8th 2022, containing the hashtags {\tt \#(I)StandWithPutin,
\#(I)StandWithRussia, 
\#(I)SupportRussia, 
\#(I)StandWithUkraine, 
\#(I)StandWithZelenskyy} and {\tt \#(I)SupportUkraine} \cite{Watt2022}. See Section \ref{s:data} for further details.
A summary plot of the data is shown in \autoref{fig:time_series_plots}. 
The figure also shows a measure of the proportion of bot traffic over the same time period, as estimated by the bot-detection tool Botometer \cite{sayyadiharikandehDetectionNovelSocial2020}.

In all time series figures, we present five significant events that provide context for our findings: when the conflict begins (24th February 2022), when the fighting in Mariupol begins (26th February 2022), when Russia captures Kherson (2nd March, 2022), when Russia captures the Zaporizhzhia nuclear power plant (4th March 2022) and when Ukrainian authorities first attempt to evacuate Mariupol (8th March 2022). These events are linked to noticeable changes in the volumes of related tweets, and in our analysis we delve deeper to understand how information is flowing.
As a result, we learn how bots are influencing the online conversation by measuring what communities are talking about online, and how this discussion evolves. We use lexicon and rule based techniques to create an approach that is robust, transferable and able to be quickly applied to large volumes of data. 

We employ time-series analysis techniques to understand how bot-like activity impacts the wider group of participants, by measuring linguistic content, sentiment and their lagged effect on future discussions. We use the Linguistic Inquiry and Word Count (LIWC; pronounced ``Luke'') \cite{liwc} and Valence Aware Dictionary for Sentiment Reasoning (VADER) \cite{vader}, dictionary based models to measure the linguistic features and sentiment of our dataset. 
To measure bot activity, we classify a random sample of 26.5\% of accounts which posted at least one English language Tweet in the dataset using Botometer \cite{sayyadiharikandehDetectionNovelSocial2020}. 

This work extends existing techniques to understand how bot-like accounts spread disinformation on Twitter and measure the effect of these malicious campaigns. The main contributions are:
\begin{itemize}
    \item An extension of existing information flow techniques to examine aggregated group activity. We establish statistical significance of information flows between accounts grouped by national lean and account type. The highest information flows are out of pro-Russian non-bot accounts. Information flows into non-bot account groups are only significant for balanced and pro-Ukraine accounts, with pro-Russian non-bot accounts only exhibiting a net outward information flow of information. 
    
    \item We establish a significant relationship between bot activity and overall non-bot sentiment (with $p = 0.000376$), but find this relationship is significant for both positive and negative lags, indicating there may be confounding factors.
    
    \item An analysis of the effect which bot activity has on emotions in online discussions around the Russia/Ukraine conflict. We find that bots significantly increase discussions of the LIWC categories: Angst, Friend, Motion, Time, Work and Filler. The strongest relationship is between Self Declared bot activity and words in the `Work' category (with $p = 3.803 \times 10^{-18}$), which includes words relating to governance structures like `president' and `government'.

    \item A dataset\footnote{Dataset available at \url{https://figshare.com/articles/dataset/Tweet_IDs_Botometer_results/20486910}.} of Twitter users who participated in discussions around the Russian Invasion of Ukraine \cite{Watt2022}. 
\end{itemize}

\section{Related work}
Many works have analysed bot-like accounts on social media \cite{orabiDetectionBotsSocial2020,kellerSocialBotsElection2019,cresciParadigmShiftSocialSpambots2017}. Authors have shown bots are present in social networks, especially with regard to political campaigns/movements \cite{orabiDetectionBotsSocial2020}. Keller and Klinger \cite{kellerSocialBotsElection2019} showed social bot activity increased from 7.1\% to 9.9\% during German election campaigns, using bot probabilities before and during the election campaign. Furthermore, Stella et al. \cite{doi:10.1073/pnas.1803470115} showed bots increase exposure to negative and inflammatory content in online social systems. These authors used various information networks to find that 19\% of overall interactions are directed from bots to humans, mainly through retweets (74\%) and mentions (25\%) \cite{doi:10.1073/pnas.1803470115}. 
A more socially-focused approach by Cresci et al. \cite{cresciParadigmShiftSocialSpambots2017}  measured Twitter’s current capabilities of detecting social spambots.
They assess human performance in discriminating between genuine accounts, social spambots, and traditional spambots through a crowd-sourcing campaign.
Notably, these works focus on analysing structural aspects of communication networks between bot and non-bot accounts, whereas we will examine information flows directly, using the full content of tweets.

Information flows in online social networks have been used to reveal underlying network dynamics, and employed to understand how individual users exert influence over one another online. Typically these flows are measured using statistical and information-theoretic measures of information flows \cite{bagrowInformationFlowReveals2019,pondComplexContagionFeatures2020,south_information_2022}, to understand if significant information flows exist between groups, particularly between bot and non-bot accounts. In social media, existing approaches only consider account-level information flows, while our work considered aggregated information flows. 

The use of bots by Russian authorities has been widely observed: {\em e.g.,} Collins
\cite{collinsMuellerReportTwitter} found 5,000 bots were pushing protests against {\em  Russiagate haux,} a political event concerning relations between politicians from US and Russia; and Shane
\cite{shaneFakeAmericansRussia2017} suggested Russia created `Fake Americans' to influence the 2016 US election. Moreover, Purtill \cite{purtillWhenItComes2022} found that Russia had a massive bot army in spreading disinformation about the Russia/Ukraine conflict. Muscat and Siebert
\cite{muscatLaptopGeneralsBot2022} have suggested that both Ukraine and Russia are utilising bot armies in their cyber warfare. However, the extent to which these bots drive particular discussions and influence the behaviour of humans on social media during the Russia/Ukraine conflict is relatively unexplored.
We aim to address this question through our analysis of information flows, sentiment, and linguistic features.

\section{Data collection and preprocessing}\label{s:data} 
We used the Twitter API (V2) to collect all tweets, retweets, quotes and replies containing case-insensitive versions of the hashtags
{\tt \#(I)StandWithPutin,
\#(I)StandWithRussia, 
\#(I)SupportRussia,} \\
{\tt \#(I)StandWithUkraine, 
\#(I)StandWithZelenskyy} and {\tt \#(I)SupportUkraine} \cite{Watt2022}. 
These Tweets were posted from February 23rd 2022 00:00:00 UTC until March 8th 2022 23:59:59 UTC, the fortnight after Russia invaded Ukraine. We queried the hashtags with and without the `I' for a total of 12 query hashtags, collecting 5,203,746 tweets. The data collected predates the beginning of the 2022 Russian invasion by one day.
These hashtags were chosen as they were found to be the most trending hashtags related to the Russia/Ukraine war which could be easily identified with a particular side in the conflict.

We first extracted all of the Twitter-labelled English tweets from the dataset. Of these, we calculated the proportion of words which appear in each LIWC category for a given tweet. These proportions are what we refer to as the `LIWC Data'. The unique accounts in this filtered data set were randomly sampled to calculate account-level Botometer labels, since Botometer uses language dependent features.

Twitter's takedown of Russian accounts on the March 3rd may lead to bias issues within our data, as the activity of these accounts will not be present in our dataset. However, analysis showed that the content spread by these accounts persisted depsite the takedown\footnote{\url{https://twitter.com/timothyjgraham/status/1500101414072520704}}.
\subsection{Categorising accounts via national lean}\label{ss:establishing_lean}

The query hashtags from each tweet were extracted and the total number of pro-Ukrainian (ending in Ukraine or Zelenskyy) and pro-Russian (ending in Russia or Putin) hashtags were counted and used to establish the national {\em lean} of a tweet. If the number of pro-Ukranian query hashtags exceeded that of the pro-Russian hashtags, the tweet was labelled as `ProUkraine', and labeled as `ProRussia' conversely. If the counts were balanced, the tweet was labelled `Balanced'. 
Where applicable, the lean of an account was taken to be the most commonly occurring national lean across all tweets from that account. 

We found that 90.16\% of accounts fell into the `ProUkraine' category, while only 6.80\% fell into the `ProRussia' category. The balanced category contained 3.04\% of accounts, showing that accounts exhibiting mixed behaviour are present in the dataset.

We explored other methods for categorising accounts, e.g., labelling accounts as `ProUkraine' or `ProRussia' if they use only those types of hashtag. However, as we were primarily concerned with aggregated activity, we elected to prioritise labelling each account by their `usual' behaviour.

\subsection{Bot Classifications} \label{ss:bot_classifications}

We use Botometer \cite{yangBotometer101Social2022}
to quantify the extent of bot activity in the dataset by assigning scores to a random sample of accounts. Note that we used Botometer's `English' scores throughout this paper -- these scores utilise both language dependent and language independent features during classification \cite{yangBotometer101Social2022}. Botometer provides an `overall' bot score, referred to as the complete automation probability (CAP) and scores corresponding to six distinct sub-types: AstroTurf, Fake Follower, Financial, Self Declared, Spammer and Other. 

The rate limits allowed us to randomly sample 26.5\% of unique accounts in our dataset which posted at least one English Tweet. This random sample leads to an approximately uniform frequency of Tweets from accounts with Botometer labels across the time frame we considered.

Due to rate limit constraints, the Botometer scores were calculated post-collection, so a small number of accounts may have been removed or scores may be calculated using activity after our collection period.

While it is more appropriate to use Botometer's CAP score as a measure of how bot-like an account is, rather than as a classification tool, it was necessary to label accounts to establish and understand information flows between account groups. 
Using the recommended cutoff of 0.43, we categorised each labelled account into one of the six Botometer categories or as `NotBot' \cite{sayyadiharikandehDetectionNovelSocial2020}. Where an account was not queried, it was labelled as `FailedToClassify'. 

The process for each account is as follows:
\begin{enumerate}
    \item If the maximum Botometer score is greater than 0.43 then the corresponding category label is assigned to that account.
     \item Else if the maximum score is smaller than 0.43, the account is categorised as`NotBot'.
     \item Otherwise the account is labelled as `FailedToClassify'.
\end{enumerate}

The results of  classification were 1,347,082 `FailedToClassify', 218,382 `NotBot', 192,633 `Other', 29,627 `Fake Follower',  29,622 `AstroTurf', 1,976 `Spammer', 1,723 `Self Declared' and 662 `Financial' accounts.

\section{The role of bots in the overall discussion} 
Figure \ref{fig:time_series_plots} shows the average hourly bot probability for different bot types (top), and the hourly frequency of query hashtags (bottom). 
There is an initial spike in the {\tt \#(I)StandWithUkraine} tweets, which is also most dominant overall. 
Interestingly, the {\tt \#(I)StandWithPutin} and {\tt \#(I)SupportPutin} hashtags spike on 2nd-3rd March, just after Russia captured its first Ukranian city (Kherson). We believe these spikes in support of Putin are predominately due to the presence of bots, as indicated by the increase in overall bot activity around this time.
This observation was independently made by researcher Timothy Graham around this time \cite{purtillWhenItComes2022}. 
On March 4th, Twitter removed over 100 users who pushed the {\tt \#(I)StandWithPutin} campaign for violating its platform manipulation and spam policy \cite{collinsTwitterBans1002022}. This may lead us to underestimate the impact of pro-Russian media after this date, as information may be spreading from alternative sources or shifting to different hashtags.

\begin{figure}[htb!]
\captionsetup{font=footnotesize}
    \begin{minipage}[b]{.5\textwidth}
        \centering
        \captionsetup{width=.95\linewidth}
    \includegraphics[width=\linewidth]{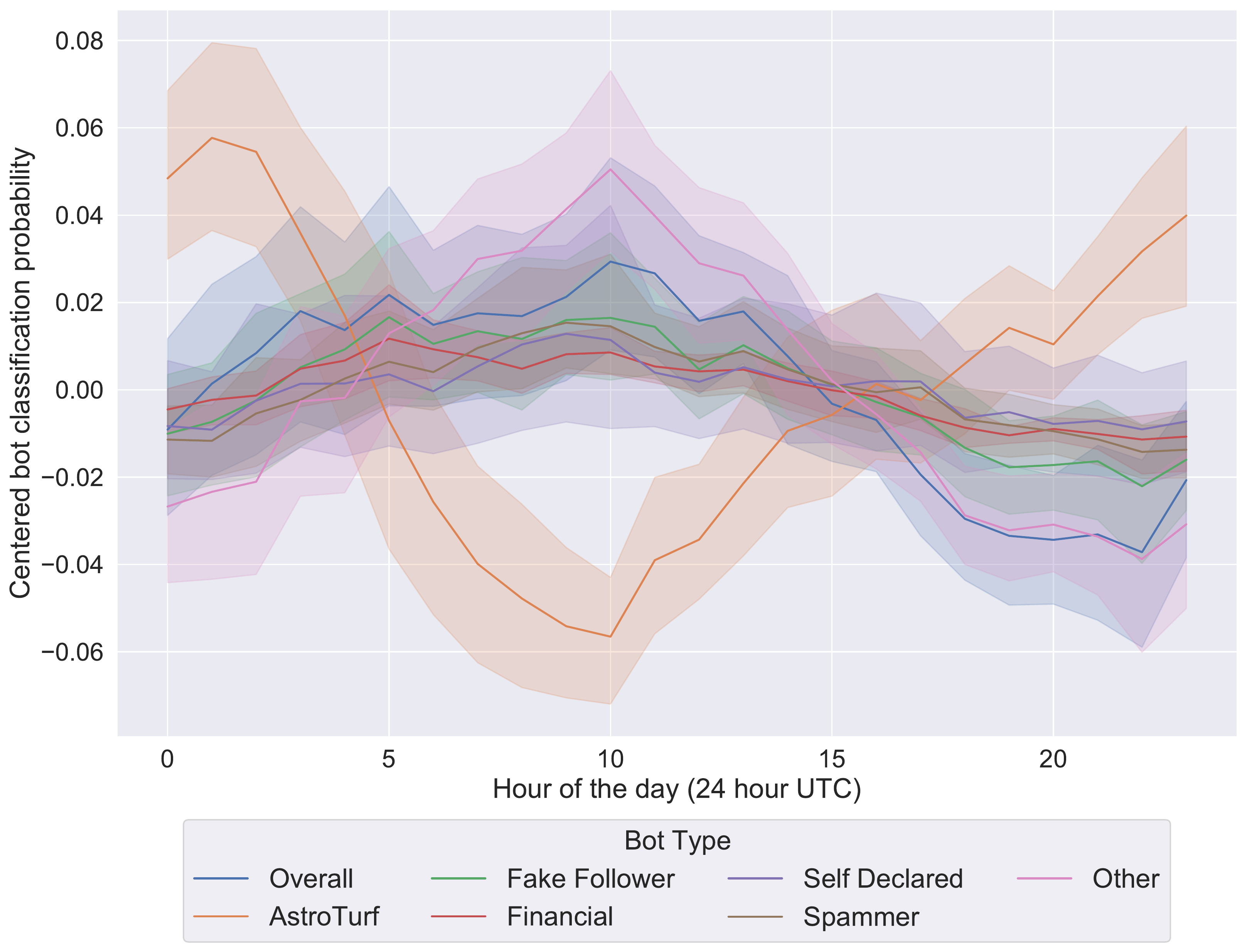}
    \vspace{-3mm}
    \caption{Average hourly Botometer results showing the daily cycle. The time series observed in Figure \ref{fig:time_series_plots} (top) is averaged based on the hour of the day (UTC time).}
    \label{fig:bot_hourly_mean}
      \end{minipage}
    \begin{minipage}[b]{.5\textwidth}
        \centering
        \captionsetup{width=.95\linewidth}
    \includegraphics[width=\linewidth]{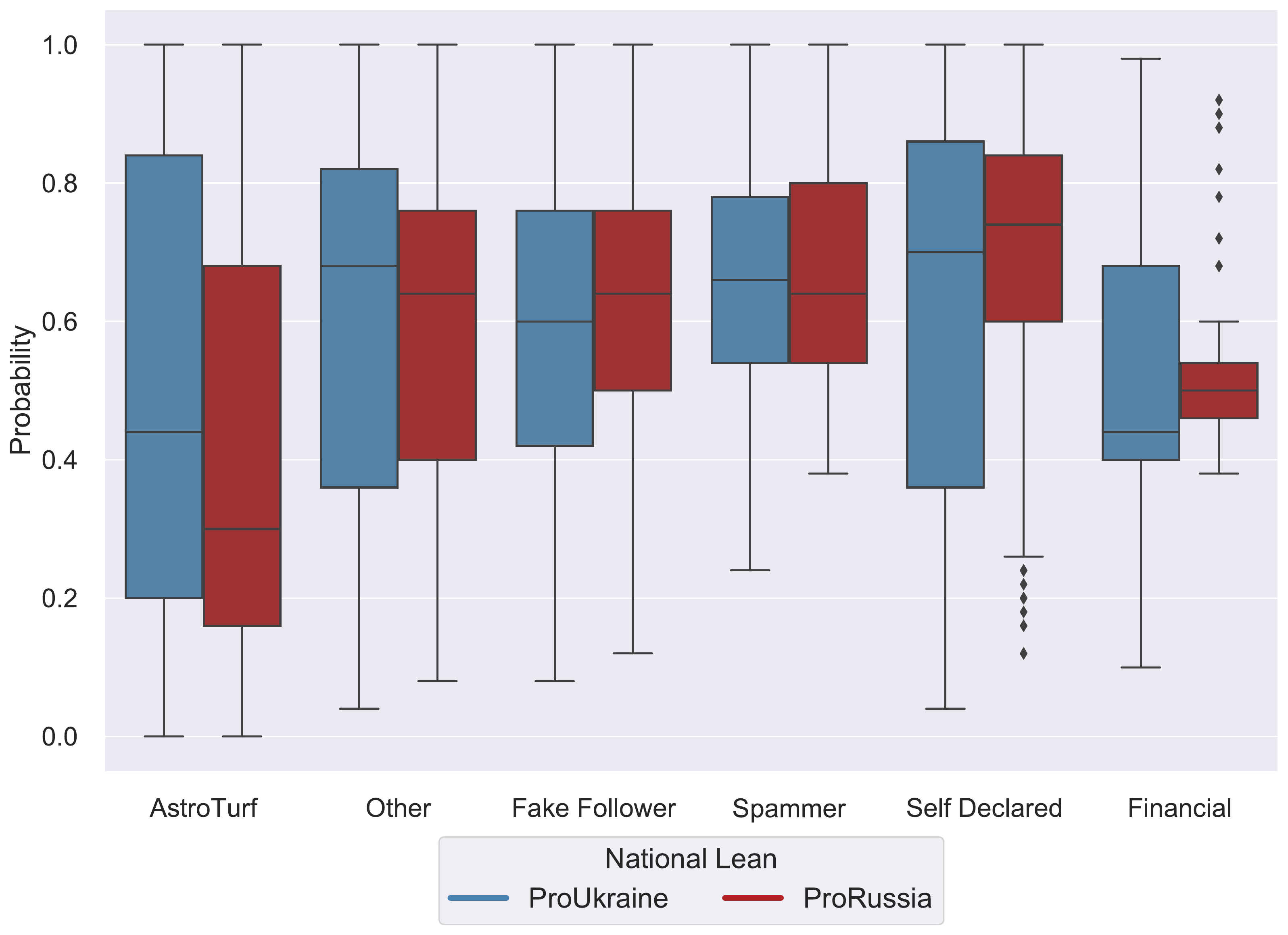}
    \vspace{0mm}
    \caption{Probabilities of bot types based on national lean and bot classification. National lean and bot classification are described in Sections \ref{ss:establishing_lean} and \ref{ss:bot_classifications}.}
    \label{fig:bot_alignment}
      \end{minipage}
\end{figure}

In Figure \ref{fig:time_series_plots} we can see the daily cycles in activity. 
 Figure \ref{fig:bot_hourly_mean} enhances that view by showing the daily cycle based on the hour of day (centred around the mean). Note that the `AstroTurf' cycle is opposite to that of all other types. Astroturfing accounts are active at opposite times to the other bot types. There are two potential explanations: either the Astroturfing accounts are from a different timezone to a majority of the accounts, or, Botometer uses timezone to determine whether an account is Astroturfing.

\vspace{-3mm}

Figure \ref{fig:time_series_plots} (top), also shows a spike in bots on March 2nd and 4th. The first spike aligns with Russia capturing Kherson, but also when the {\tt \#(I)StandWithPutin} and {\tt \#(I)StandWithRussia} hashtags were trending. We observed the mean overall Botometer score of active pro-Russia accounts increased from 0.535 (1st March) to 0.593 (2nd March), whereas the mean overall Botometer score of active pro-Ukrainian accounts decreased from 0.585 (1st March) to 0.562 (2nd March). Hence, this further suggests that bots were responsible for making the pro-Russia hashtags trend on these dates.
The second spike in bot activity on March 4th is more difficult to explain. On this date Russia captured the Zaporizhzhia nuclear power plant, but also a handful of pro-Russian accounts were removed by Twitter. The mean overall Botometer score of active pro-Russia accounts significantly increased from 0.535 (3rd March) to 0.613 (4th March) and the mean overall Botometer score of active pro-Ukrainian accounts slightly increased from 0.573 (3rd March) to 0.603 (4th March).
As a result, this spike in bot activity is due to the presence of pro-Russian bots (before they were removed) and the presence of pro-Ukrainian bots advocating against the pro-Russian accounts. 
Nonetheless, there is an obvious presence of bots over the duration of the first fortnight after Russia's invasion of Ukraine.

The time of day effects are most pronounced for the AstroTurf and Other bots, whereas the activity of Fake Follower, Financial, Self Declared and Spammer bots are less impacted by the time of day. 
This may be because AstroTurf and Other bots are pushing campaigns specific to certain countries, and hence sharing content aligned with those timezones. 
The spike in Other bots occurs at 10:00 UTC which corresponds to 1:00pm Ukrainian time. 
Matthews \cite{matthewsBestTimeTweet2015} suggested that noon to 1:00pm is the most popular time to tweet in any timezone. Hence, the Other bots are likely to be increasing their engagement in Ukraine by being most active around this time.

Figure \ref{fig:bot_alignment} shows pairwise box plots of the Botometer type probabilities based on whether the accounts are pro-Ukraine or pro-Russia. 
The most commonly-used bot type for both campaigns is the Self-Declared bots,
suggesting that authorities have identified these bots to be most useful in a information warfare campaign. 
Furthermore, we observe a fairly consistent spread of bot types for both campaigns. Pro-Russian accounts have a mean CAP score of 0.42, while pro-Ukrainian accounts have a mean score of 0.43, with medians 0.36 and 0.34 respectively. 
However, the median probability of an account being an AstroTurf bot is slightly higher for pro-Ukrainian accounts than pro-Russian accounts. 
Additionally, the median probability of a Self-Declared bot is slightly higher for pro-Russian accounts compared to pro-Ukrainian accounts. 
This highlights that pro-Ukrainian accounts may be utilising more Astroturfing in their information warfare, whereas pro-Russian accounts may be utilising more Self-Declared bots.

\section{Information flows between bots and human accounts}

\subsection{Information-flow estimation methods} \label{s:methods}
We measure the influence of accounts on overall online discussion using the following symmetric net information flow measure from the time-stamped writings of a source $\mathcal{S}$ to target $\mathcal{T}$ \cite{south_information_2022}: 
\begin{equation}
    \label{tobineq}
    \Delta(\mathcal{T} || \mathcal{S})= \dfrac{\hat{h}(\mathcal{T} || \mathcal{S})}{\sum_X \hat{h}{(\mathcal{T} || X)}} - \dfrac{\hat{h}(\mathcal{S} || \mathcal{T})}{\sum_X \hat{h}{(\mathcal{S} || X)}}.
\end{equation}
Here $\hat{h}(\mathcal{T}||\mathcal{S})$ is the non-parametric cross entropy rate estimator \cite{bagrowInformationFlowReveals2019,kontoyiannis1998nonparametric}:
\begin{equation}
\hat{h}(\mathcal{T} || \mathcal{S})=\frac{N_{\mathcal{T}} \log_{2} N_{\mathcal{S}}}{\sum_{i=1}^{N_{\mathcal{T}}}\Lambda_{i}(\mathcal{T}| \mathcal{S}_{\leq t(T_i)})},
\label{eqn:cross}
\end{equation}
where $N_S$ and $N_T$ are the number of symbols written by the source and target, respectively, and $\Lambda_i^i$ denotes the length of the shortest substring, $l$ starting at index $i$ which does not appear in the first $i+l-1$ symbols.
See \cite{bagrow2018quoter} for an example of $\Lambda_i^i$ estimation.
We aggregate content by account type rather than on an account level to measure the information flows between account types and establish their significance.

We use the language analysis tools Valence Aware Dictionary and Sentiment Reasoner (VADER) \cite{vader} for sentiment analysis, as well as the Linguistic Inquiry and Word Count (LIWC) \cite{liwc} to establish relationships between conversation features and bot-activity. 
We then use the Granger causality test to determine whether one time series $X$ is useful in forecasting another time series $Y$ with some time lag $p$.

We do this by fitting two linear models. The first model we include only the lagged values of $Y$:
\begin{equation}\label{granger_eq1}
    Y_t = \alpha_{1,0} + \alpha_{1,1} Y_{t-1} + \dots + \alpha_{1,p} Y_{t-p} + \epsilon_{1,t},
\end{equation}
where we define $\epsilon_{i,t}$ as the error term of model $i$ at time $t$ and $\alpha_{i,j}$ as the parameter of model $i$ at lag $j$. Next, we augment the model to also include the lagged values of $X$:
\begin{equation}\label{granger_eq2}
    Y_t = \alpha_{2,0} + \alpha_{2,1} Y_{t-1} + \dots + \alpha_{2,p} Y_{t-p} + \beta_{1} X_{t-1} + \dots + \beta_{p} X_{t-p} + \epsilon_{2,t}.
\end{equation}
The null hypothesis, that $X$ does not Granger-cause $Y$, is accepted via an F-test if and only if no lagged values of $X$ are retained in the regression model observed in Eq. \ref{granger_eq2}.
\subsection{Aggregated information flows}
We apply information-flow measures to content aggregated by account type, to understand inter-community information flows. Rather than using an aggregate statistic on individual information flows, the proposed aggregated flow approach allows the symmetric and normalisation properties of the net information flow measure \cite{south_information_2022} to be preserved. This process improves the quality of the entropy estimate for the group behaviour, by increasing the available sequence length and mitigating the effect of slow convergence of the estimator. In this section we also develop a significance test for net information flow.

Each account is labeled by both bot classification and national lean, and the content within these account groups is aggregated. 
The cross entropy between each of these groups is calculated pairwise, and these values are normalised according to Eq. \ref{tobineq}.

\begin{figure}[htb!]
\captionsetup{font=footnotesize}
    \includegraphics[width=\linewidth]{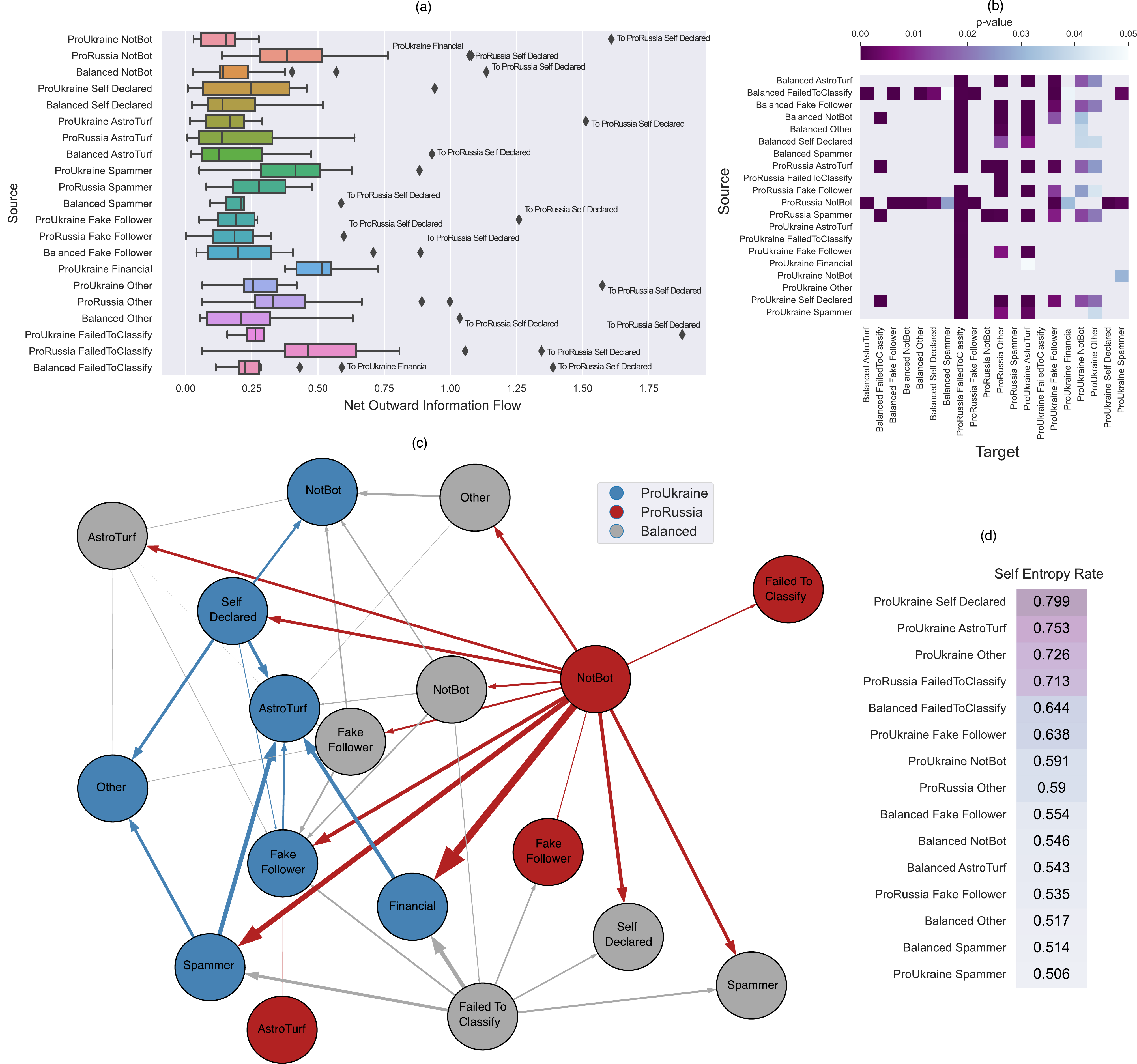}
    \caption{(a) Aggregated net outward information flows by account type and national lean. The outward flows for each aggregated group predominantly fall in the $(0,0.5)$ range, however `ProRussia NonBot', `ProUkraine Financial', `ProUkraine Spammer' and `ProRussia FailedToClassify' net flows tend to be greater. Values above the 80th percentile are labelled -- the majority of these represent flows into the `ProRussia Self Declared' group. (b) Heatmap of empirical $p$-values for each intergroup net information flow, showing significance of the difference in median outbound flows between groups. This test is used to form the network shown in (c). (c) Significant net information flows between groups. Not pictured are aggregated groups with inadequate sample size. Each aggregated group is coloured by national lean; edges are weighted by magnitude of net information flow. Significant flows out of `ProRussia NotBot' accounts indicate that information flows from these accounts to other groups, most commonly groups with a balanced or `ProUkraine' lean. Most intergroup flows for groups with a `ProUkraine' lean are to other groups with `ProUkraine' lean, with no significant flows from `ProUkraine' account groups to `ProRussia' account groups. (d) Lists the 15 highest self entropy rate values for the aggregated groups, with the top three groups all having a `ProUkraine' lean.  \label{fig:infofig} \vspace{-5mm}}
\end{figure}

These pairwise cross entropy estimations produce a fully connected network. We then perform a statistical test for whether aggregated net information flows are significant between groups, allowing a network of significant information flows to be constructed (Figure \ref{fig:infofig}).

To approximate the null distribution for the difference between median outgoing flow rates between each group, we randomly shuffle group labels for each tweet, reconstruct aggregated sequences, then calculate net information flows. 
These aggregated net flows are used to calculate the differences in group median out-flows and construct the empirical null distribution, which is used to calculate an empirical p-value for the observed values.
These aggregated net information flows reveal that generally information flows out of the pro-Russian accounts, with the exception of the  pro-Russia FailedToClassify and  pro-Russian Fake Follower account groups (Figure \ref{fig:infofig}). This indicates that these account groups may be predominately interacting with other accounts within the same group, rather than accounts with other leans or types. The `ProRussia NotBot' account group has the largest outward information flows and significant flows to a range of other groups, having a positive information flow into both `ProUkraine' and `Balanced' account groups.

This indicates that these Russian non-bot accounts influence a variety of user groups with the greatest between group information flows. This may indicate that human-controlled accounts, or accounts which appear less bot-like, have more influence in our social network, potentially due to their behaviour or perception. While the `NotBot' label is derived from the Botometer score, this label does not mean these accounts are not malicious or automated.

Most of the significant information flows between `ProUkraine' account groups is between groups with the same lean. This may indicating that more information flows between the accounts within each of these groups rather than to accounts in other groups. `ProUkraine' groups have the highest self entropy rates, meaning that these groups do not just aggregate information from other account groups, but influence other accounts within the same group (Figure \ref{fig:infofig} (d)).

The Balanced account groups show information flows to all other national-lean types, and connect otherwise disjoint parts of the information flow network. These accounts may act as a bridge for information to move between `ProRussia' and `ProUkraine' accounts. Most of these groups have small but significant information flows to other groups, with information tending to flow out of these groups.

Notably, the few significant information flows into non-bot account groups indicate  some influence from `bot-like' accounts on non-bot accounts. However, these account groups have stronger outward net information flows then inward flows, suggesting that while they tend to have influence on the content of other `bot-like' accounts, they do not influence non-`bot-like' users generally. 

When account-level flows are considered rather than the aggregated flows presented here, several similar significant flows exist between `bot-like' and non-`bot-like' accounts.

\section{How bot accounts influence linguistic features of the conversation} 
Having characterised bot activity and identified significant information flows, we now aim to explore the content of these relationships.
We first consider relationships between bot activity and sentiment, with a focus on understanding if bot-like accounts have a significant impact on the compound sentiment of non-bot accounts, measured using the CAP Botometer score and weighted average compound sentiment. 
The linguistic impact is then quantified by using LIWC to develop a statistical framework for understanding the relationship between bots activity and emotional/linguistic content.

\subsection{Bot activity and overall sentiment}
To understand how bots drive non-bot sentiment, we begin by cleaning and preparing two time series. 
The first is the mean CAP Botometer score, which acts as a proxy for the total proportion of bot-like activity on the network. The second is the CAP-weighted mean compound sentiment.
Weighting the VADER compound sentiment by the complement of the Botometer CAP score provides a measure of non-bot sentiment without making account labelling assumptions. 
It is robust to threshold choices, and provides a meaningful measure of the overall sentiment of the dataset. 
\begin{figure}[t]
\captionsetup{font=footnotesize}
    \includegraphics[width=\linewidth]{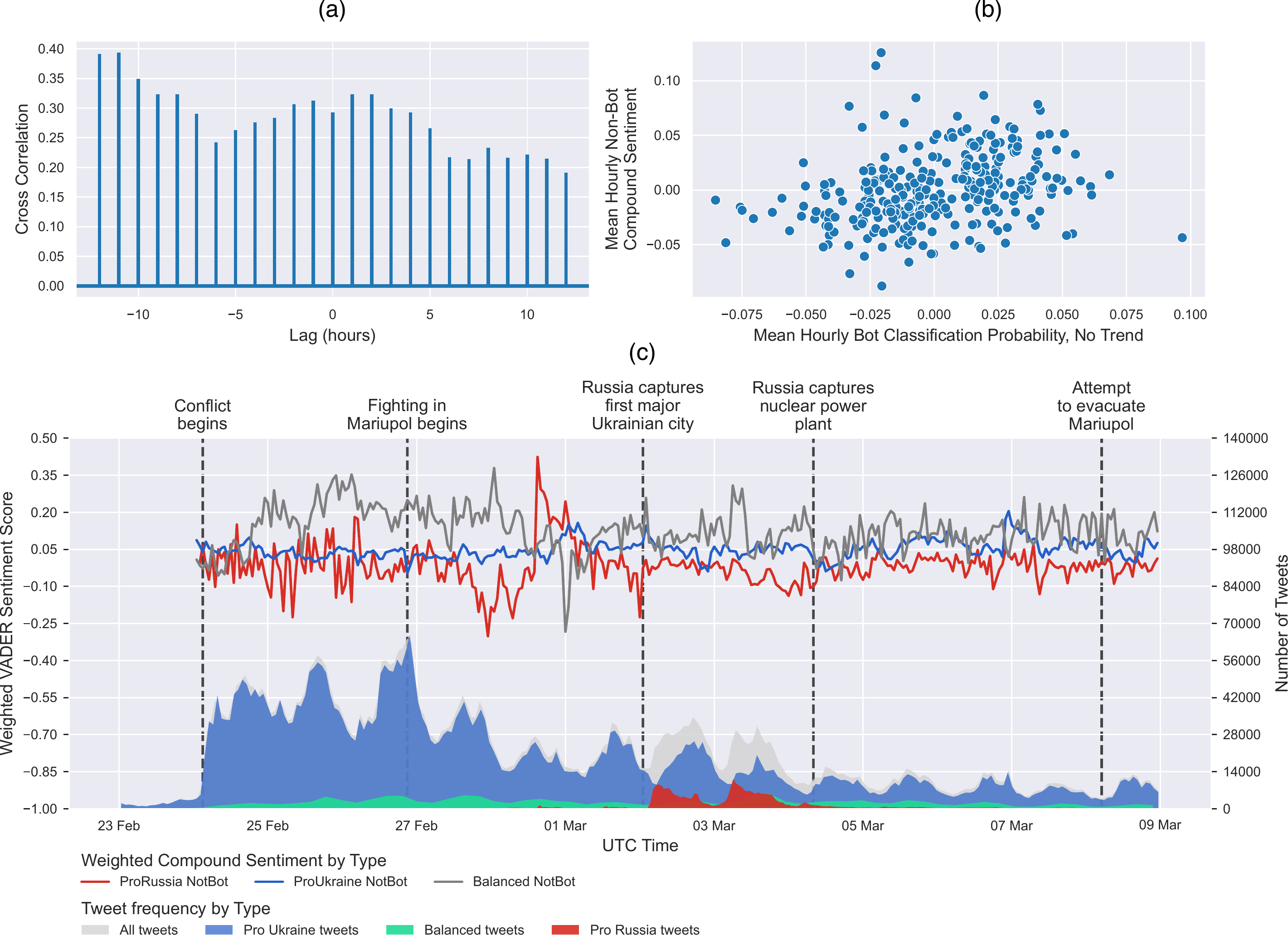}
    \caption{(a) Shows the lagged cross correlations between the mean hourly Botometer Overall CAP score and the CAP weighted mean compound sentiment. Each lag represents an offset of the CAP weighted mean compound sentiment in hours. These correlations are significant in both lag directions. (b) Considering a scatterplot of the two timeseries with trend and burn-in samples removed, reveals a positive linear relationship between them. (c) The timeseries represent the CAP weighted mean compound sentiment grouped by national lean, with `ProRussia', `ProUkraine' and `Balanced' accounts considered separately. To aid interpretation of these timeseries, the tweet frequency of each type (from all accounts) and some significant event markers are given. Before the 2nd of March there was minimal activity from `ProRussia' accounts, so the CAP weighted mean compound sentiment estimate has high variance and is of low quality. After March 3rd, there is a spike in Balanced CAP weighted mean compound sentiment, from Balanced accounts, suggesting that these accounts were producing more positive tweets overall, potentially in response to humanitarian corridors opening. \vspace{-4mm}
    \label{fig:sentifig}}
\end{figure}

Each time series is aggregated hourly. The first 50 hours are removed from both time series since there is a comparatively small tweet volume over that period.The mean CAP Botometer score has a linear trend, which is removed via a linear regression. 
Both time series are standardised to have mean zero to ensure they comply with the assumptions to perform Granger causality analysis.
We also removed the daily periodic cycle (Figure \ref{fig:bot_hourly_mean}) from each time series.

The cross correlations are then calculated for various lags to understand the effect of mean CAP Botometer score on the CAP-weighted mean compound sentiment. A maximum lag of 12 hours is considered.
A positive relationship exists between the cleaned time series, indicating there is a correlation between the activity of `bot-like' accounts and the compound sentiment of the non-`bot-like' accounts. There is a significant relationship between the two series, with $p = 3.76 \times 10^{-4}$. 

Since effects cannot occur simultaneously, we consider the lagged effect of bot activity on non-bot compound sentiment, finding a positive cross correlation for both positive and negative lags (Figure \ref{fig:sentifig}). This shows that bot activity increases when sentiment increases, but also that sentiment increases with increases in bot activity. Figure \ref{fig:sentifig} indicates that the relationship between sentiment and bot activity is complicated, with marked events driving spikes in the compound sentiment of not-`bot-like' accounts. There are also spikes in the mean compound sentiment of other account lean types, which may be due to events which we did not consider in our analysis.
This indicates that there may not be an overall effect on the non-bot compound sentiment due to bot activity, although this relationship may exist on an individual account level.
\subsection{Bot activity and linguistic discussion features} 
Using LIWC, we explore how different types of bots drive emotions and discussions around the Russia/Ukraine conflict. 
We produce hourly averages for overall LIWC proportions and the Botometer probabilities. This results in a set of time series, all over 336 hours. We utilise the Granger Causality Test (Section \ref{s:methods}) on these time series to determine whether the activity of certain bots Granger-cause more or less discussion around particular LIWC categories.

We apply pairwise Granger Causality Tests between each Botometer timeseries, $X$, and each LIWC timeseries, $Y$, for $p = 12$ lags/hours (see Eq. \ref{granger_eq1} and \ref{granger_eq2} in Section \ref{s:methods}). This time window is chosen as it is reasonable to assume a majority of the effects from bots will occur over this time frame. The validity of this assumption is explored below. 

\begin{figure}[htb!]
\captionsetup{font=footnotesize}
\begin{centering}
    \includegraphics[width=0.9\linewidth]{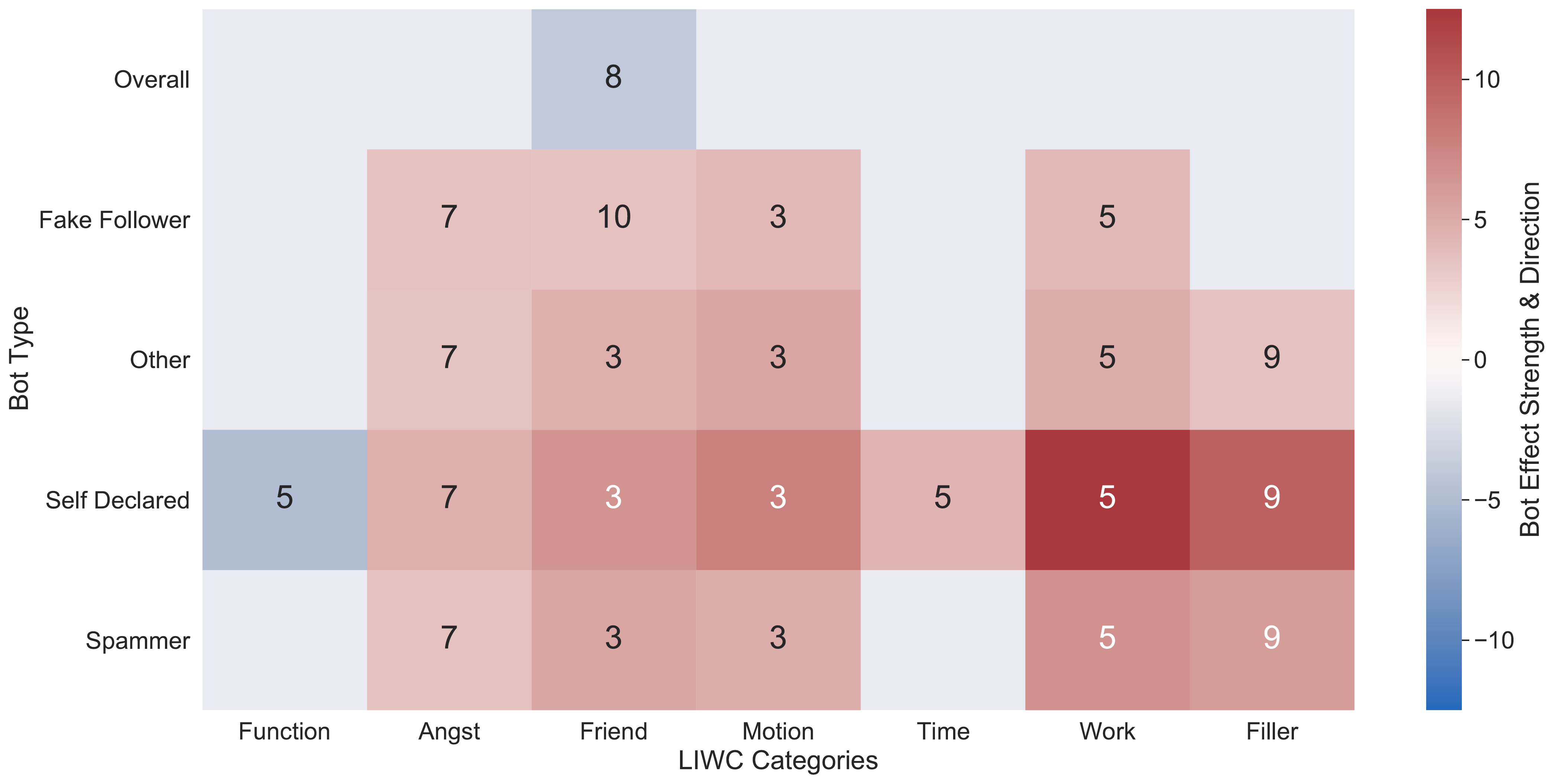}
    \caption{A series of pairwise Granger Causality Tests are performed to examine whether the activity of bot types is Granger-causing changes in discussions of the LIWC categories. The heat maps colour describes the bot effect strength and direction from the Granger Causality Test (over 12 hours/lags) between the time series of hourly bot proportions and the time series of hourly LIWC category proportions. The number in the centre describes the most prolific lag in the Granger Causality Test. We calculate the bot effect strength using the F-score from an F-test on the Granger Causality linear models. Moreover, we calculate the bot effect direction and most prolific lag using the sign and lag (respectively) of the largest $\beta$ coefficient from Eq. \ref{granger_eq2} in Section \ref{s:methods}. We perform a Bonferroni Adjustment on the p-values from the Granger Causality Tests and only show the Bot Types and LIWC Categories with a significant adjusted p-value $(< 0.05)$.\vspace{-5mm}}
    \label{fig:bot_granger_liwc}
\end{centering}
\end{figure}

\begin{figure}[htb!]
\captionsetup{font=footnotesize}
    \centering
    \begin{subfigure}{.32\textwidth}
      \centering
      \includegraphics[width=0.9\linewidth]{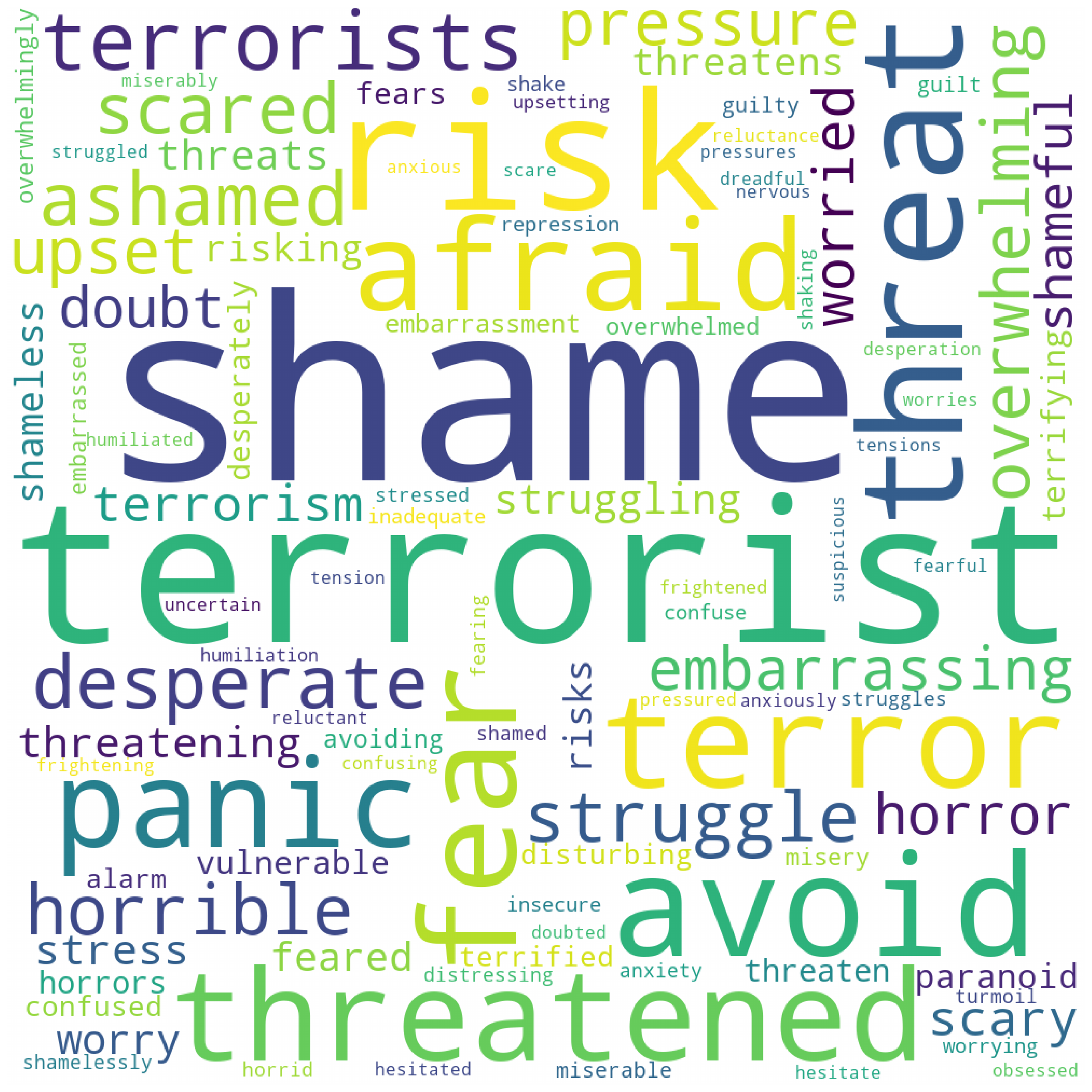}
      \caption{Angst Category}
      \label{fig:anx_cloud}
    \end{subfigure}
    \begin{subfigure}{.32\textwidth}
        \centering
        \includegraphics[width=0.9\linewidth]{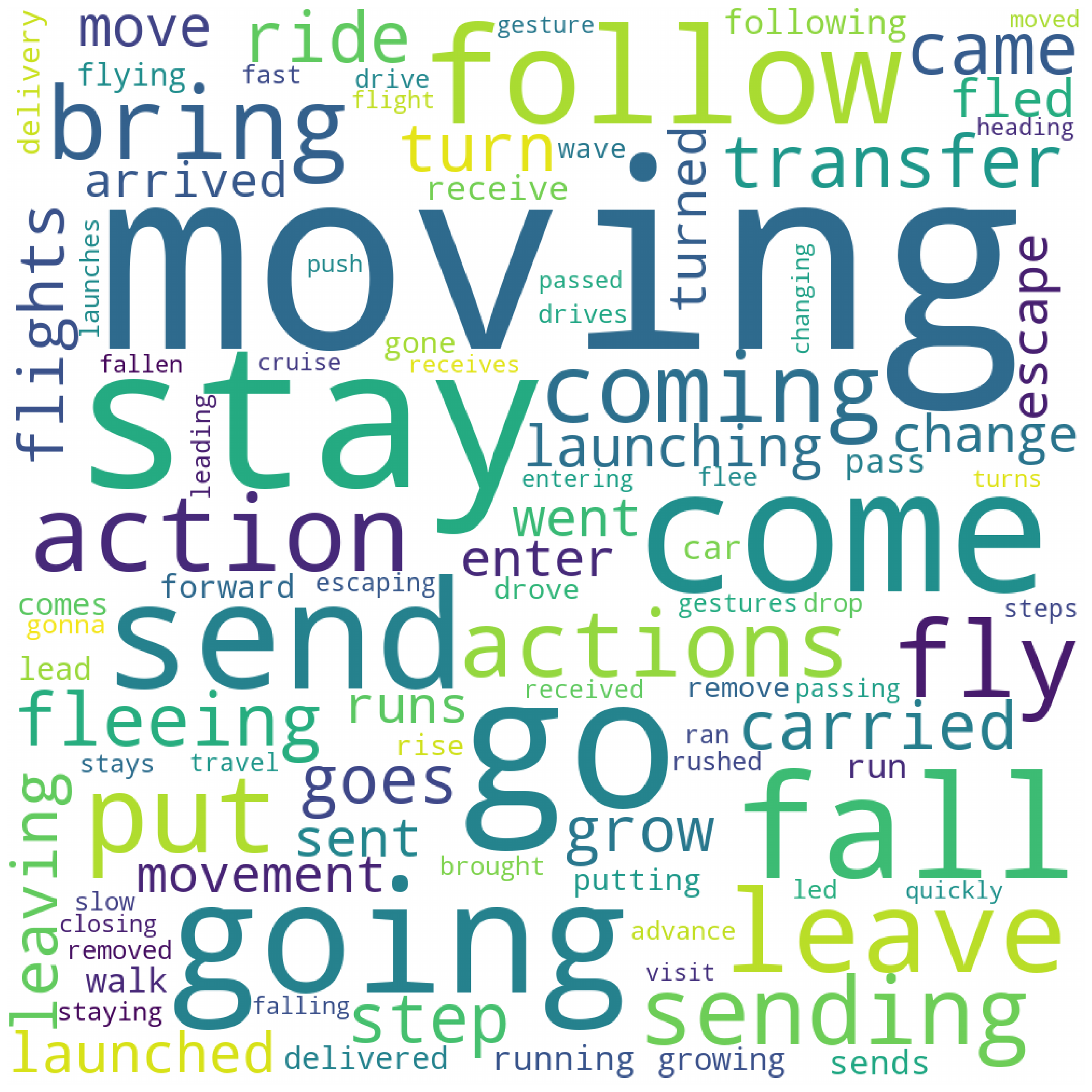}
        \caption{Motion Category}
        \label{fig:motion_cloud}
      \end{subfigure}
    \begin{subfigure}{.32\textwidth}
      \centering
      \includegraphics[width=0.9\linewidth]{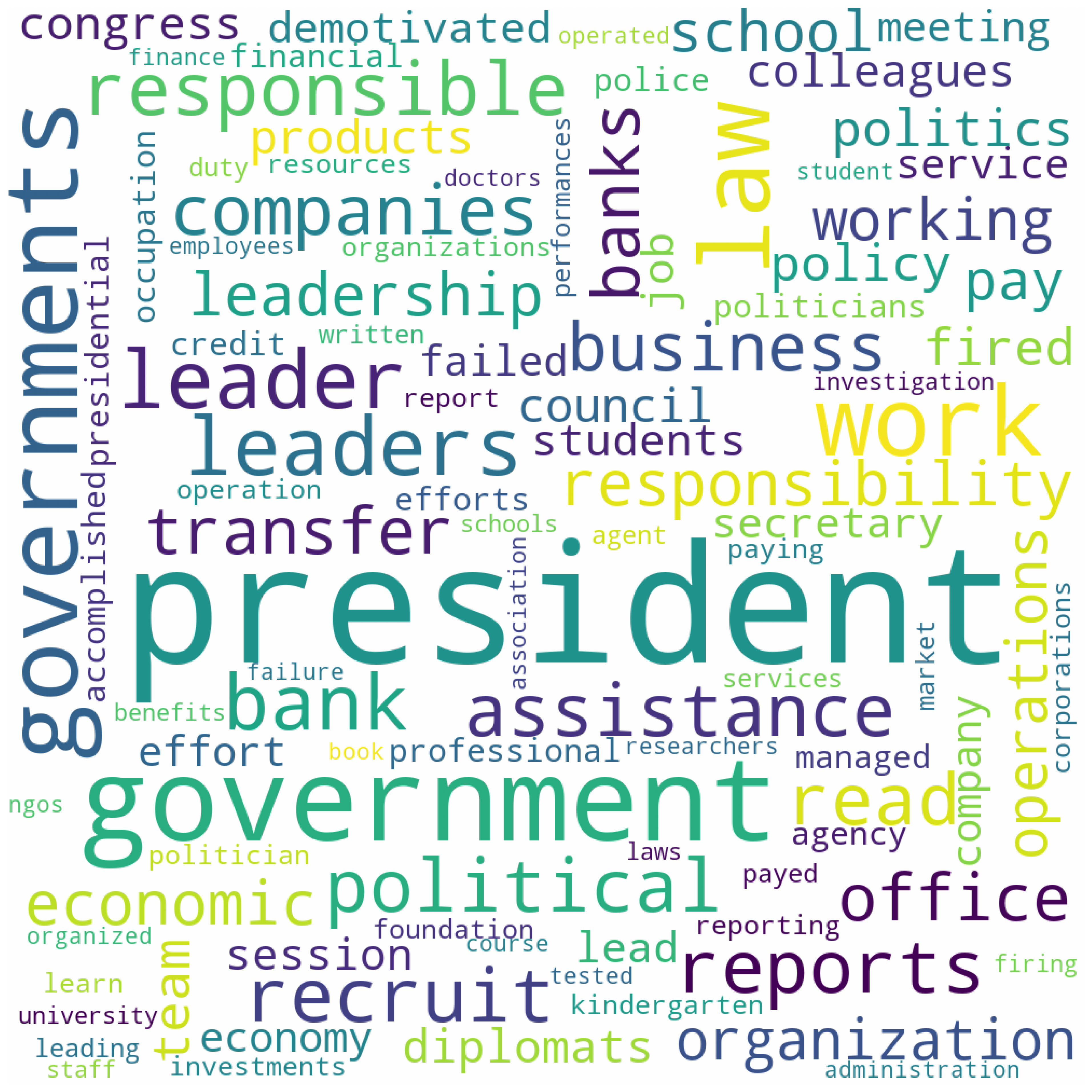}
      \caption{Work Category}
      \label{fig:work_cloud}
    \end{subfigure}
    \caption{Word Clouds which demonstrate the frequency of words in particular LIWC categories.\vspace{-5mm}}
    \label{fig:word_clouds}
\end{figure}

We use the F-score from the Granger Causality Test as a measure of how `influential' a type of bot is on each LIWC category. 
To get a sense of direction for these relationships, we use the sign of the largest $\beta$ coefficient from Eq. \ref{granger_eq2} in Section \ref{s:methods}. 
We multiply the sign of this coefficient by the F-score from the Granger Causality Test to obtain a measure of strength and direction, referring to this as the Bot Effect Strength and Direction. Moreover, we use the lag of the largest $\beta$ coefficient from Eq. \ref{granger_eq2} in Section \ref{s:methods} to represent the most prolific lag in the relationship. We use the p-value from the Granger Causality Test to determine whether the effects are significant, and perform a Bonferroni Adjustment to adjust for multiple hypothesis tests. The results are displayed in Figure \ref{fig:bot_granger_liwc}, where we have only included the significant relationships. The number in the centre of each square represents the most prolific lag -- we interpret this as the number of hours until the effects of the bot activity are most pronounced.
\vspace{-3mm}
Figure \ref{fig:bot_granger_liwc} shows that bots do have a significant impact on discussions of certain LIWC categories. 
To better understand what each of these LIWC categories represent, we generated word clouds of the words from each LIWC category that appeared in the dataset. 
The size of the words represent their relative frequency in the data -- the larger the word the more frequently it occurs. The word clouds for Angst, Motion and Work are shown in Figures \ref{fig:anx_cloud}, \ref{fig:motion_cloud} and \ref{fig:work_cloud}, respectively. 
For a full discussion of the words associated with various LIWC categories, see Pennebaker and Francis \cite{pennebakerCognitiveEmotionalLanguage1996}.
In Figure \ref{fig:bot_granger_liwc}, the self-declared bots have greatest amount of influence on a number of discussions.
In particular, the self-declared bots increase discussions around angst, friends, motion, time, work and the usage of filler words, but decrease the usage of function words. 
Moreover, it is apparent that these bots most strongly influence discussion of the work category (with a most prolific lag of five hours). 
Figure \ref{fig:work_cloud} shows that most of the discussion around work is involved with governing bodies, with `president' and `governments' being the most commonly used words. 
While it is difficult to assert exactly why these bots are driving more discussions of work, we gain further understanding by also observing that self declared bots Granger-cause more angst (with a most prolific lag after 7 hours). 
Combining these two observations suggests that self-declared bots drive more angst about governing bodies. 
From a pro-Russian perspective, this may be to cause more disruption in the West, and from a pro-Ukrainian perspective, this may be to cause more disruption in Russia.
Figure \ref{fig:bot_alignment} shows a fairly even probability of pro-Russia and pro-Ukraine accounts being self-declared bots. 
Although the exact origin of self-declared accounts is unknown, it is worth noting that we considered predominately English accounts. 
It is therefore more likely that the intention of these accounts was to drive more disruption in English-speaking countries.

\begin{figure}[b!]
\captionsetup{font=footnotesize}
    \includegraphics[width=\linewidth]{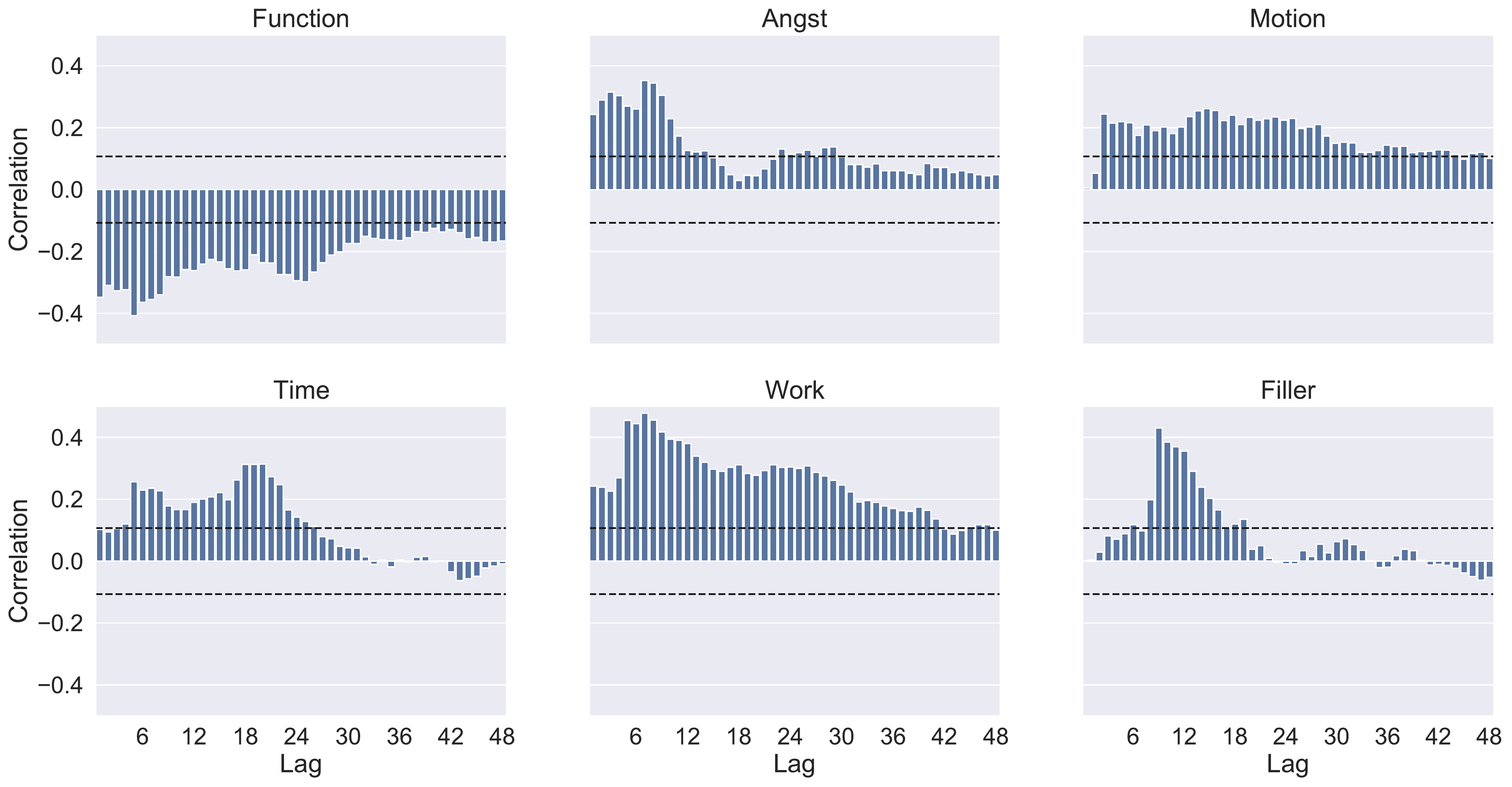}
    \caption{Lagged cross correlations between the hourly Self Declared bot proportions and the significant hourly LIWC Category proportions (significance is determined from the results in Figure \ref{fig:bot_granger_liwc}). We consider 48 hours/lags for each of these plots and represent the significance threshold using a horizontal dotted line. This shows the extent to which the bots drive changes in online discussion and how long these effects can persist for. \vspace{-5mm}}
    \label{fig:cross_correlations_liwc}
\end{figure}

Observe that Fake Follower, Spammer and Other bots also increase angst (all with the most prolific lag after 7 hours). 
Figure \ref{fig:anx_cloud} shows that a majority of angst-related words are surrounding fear and worry. 
Hence, we argue that self-declared, fake follower, spammer and other forms of automated account types combine to increase fear in the overall discussion of the Russia/Ukraine war. 
This observation has been hypothesised by many authors \cite{nguyenHowPutinPropaganda2022,osborneUkraineDestroysFive2022}, but a detailed analysis has been lacking and may be of concern for many governments and defence organisations.

Figure \ref{fig:bot_granger_liwc} further shows that fake follower, self-declared, spammer and other bot types also increase online discussion around motion. In Figure \ref{fig:motion_cloud}, we see a number of motion related words that are potentially associated with staying or fleeing the country. 
Combining this with increases in Angst suggests that bots could be influencing people's decisions surrounding whether to flee their homes or not. Druziuk \cite{druziukCitizenlikeChatbotAllows2022} noted that bots have allowed ``Ukrainians to report to the government when they spot Russian troops'', but the usage of bots to influence people on staying/leaving the country is something not observed before. However, it is difficult to denote whether this is being done in support of Ukraine, Russia or both. 

In Figure \ref{fig:bot_granger_liwc} the most prolific lag is mostly consistent for a given LIWC category, but varies greatly for bot type. Hence, the time which bots effect a given discussion on the war depends mainly on the topic of discussion and not on the type of bot. For instance, we observe that Fake Follower, Self-Declared, Spammer and Other bots all most prolifically effect discussions of work after five hours. To further examine the effects of the lag on discussion of different LIWC categories, we plot cross correlations in Figure \ref{fig:cross_correlations_liwc}. These plots represent the cross correlation between Self-Declared bot proportions and a number of significant LIWC categories (in Figure \ref{fig:bot_granger_liwc}) over 48 hours.

The direction of the effect for each LIWC variable in Figure~\ref{fig:cross_correlations_liwc} is consistent with Figure~\ref{fig:bot_granger_liwc}, further validating our results. 
This direction is consistent for all significant lags, justifying our decision to choose the largest parameter in the regression model as an indication of direction in Figure~\ref{fig:bot_granger_liwc}. 
For some LIWC categories the effects of Self-Declared bots linger over many lags but for others the effects diminish relatively quickly. For instance, the effects on the work category and the function category are significant for lags almost up to 48 hours, whereas the effects on the angst and filler categories diminish within 24 hours. 

\section{Conclusion}

This work investigates if and how bot-like accounts influence the online conversation around the Russian invasion of Ukraine during February 2022.
We showed which account groups have measurable influence by aggregating accounts using their national lean and account bot-type label. The patterns of information flows between bot and non-bot account vary based on national lean: Pro-Russian non-bot accounts are most influential overall, with information flows to a variety of other account groups. No significant outward flows exist from pro-Ukrainian non-bot accounts, with significant flows from pro-Ukrainian bot accounts into pro-Ukrainian non-bot accounts. Pro-Russian account groups are seemingly isolated, with smaller self-entropy rates and less significant between group net information flows. However, there exists significant information flows out of pro-Russia non-bot and AstroTurf account groups, with the largest net flows originating in the pro-Russian non-bot account groups. Contrastingly, pro-Ukrainian account groups tend to have more information flows between pro-Ukrainian and balanced account groups. Pro-Ukrainian aggregated groups also tend to have higher self-entropy rates. 

To understand how bot-like accounts influence non-bot like accounts across all national lean types, we consider non-bot sentiment, measured using a weighted compound sentiment score. By weighting this by the overall bot probability, this compound sentiment reflects the overall sentiment of non-bot-like accounts across the network. The relationship between this sentiment and the bot activity is significant but occurs in both directions, with sentiment and bot activity both impacting each other.

Finally, we identify the effect of bot-like accounts on LIWC linguistic discussion features. 
Self-declared bots have the largest impact, showing significant relationships with word in the Function, Angst, Friend, Motion, Time, Work and Filler categories. To find the direction and significance of these relationships we use pairwise Granger causality tests. We find bots generally increase word usage in these categories with a 3-10 hour lag. Self-declared 
bots show the strongest relationship with Work discussions, i.e., words pertaining to governing bodies (``president'', ``government'' and ``leadership''). We also see bot accounts increase the use of words in the 
angst category which contains words related to fear and worry (``shame'', ``terrorist'', ``threat'', ``panic'').

In future work, we will explore information contained on the network of interactions between users recorded in the tweets using a network science approach \cite{50_barabasi2016network,newman2018networks,caldarelli2007scale}. We will also further explore diverse ways to classify the national lean of authors based on their published Twitter content. Preliminary results indicate heavy-tailed distributions in timing lags that differ between account types, suggesting differences in coordinated activity signatures \cite{mathews2017nature}.
We will examine coordination campaigns and coordination networks~\cite{schoch2022coordination,giglietto2020takes,pacheco2021uncovering,keller2020political,lukito2020coordinating}  to quantify the impact of coordinated activity in the social network structure and to further investigate its influence on social media users.

Using a number of approaches we describe a framework to understand the impact of bots on the network, explore how malicious campaigns operate and measure their effect on online discussion. Our approach is applicable to any social media content presenting polarisation between distinct groups and can be applied to other data to understand how malicious campaigns operate.

\section{Acknowledgements}

B.S. would like to acknowledge the support of a Westpac Future Leaders Scholarship.
L.M. and M.R. are supported by the Australian Government through the Australian Research Council’s Discovery Projects funding scheme (project DP210103700).
L.M. also acknowledges support from the Australian Defence Science and Technology Group ORNet scheme.

\newpage
\bibliographystyle{splncs04}
\bibliography{References1} 


\end{document}